\documentclass[twocolumn,prl,showpacs,preprintnumbers,amsmath,amssymb]{revtex4}

\usepackage{graphicx}
\usepackage{dcolumn}
\usepackage{bm}

\begin{document}

\title{Topological defects 
in antiferromagnetically coupled 
multilayers with perpendicular anisotropy
}
\author{N. S. Kiselev$^{1,2}$}
\thanks
{Corresponding author 
}
\email{kiselev@kinetic.ac.donetsk.ua}
\author{U. K. R\"o\ss ler$^{1}$}
\email{u.roessler@ifw-dresden.de}
\author{ A. N.\ Bogdanov$^{1}$}
\author{O. Hellwig$^{3}$}

\affiliation{$^1$IFW Dresden, Postfach 270116, D-01171 Dresden, Germany}%
\affiliation{$^2$Donetsk Institute for Physics and Technology, 
83114 Donetsk, Ukraine}
\affiliation{$^3$San Jose Research Center, Hitachi Global Storage Technologies,
 San Jose, CA 95135, USA}
\begin{abstract}
{A rich variety of specific multidomain textures
recently observed in antiferromagnetically coupled
multilayers with perpendicular anisotropy include
regular (equilibrium) multidomain states as well
as different types of \textit{topological}
magnetic defects. 
Within a phenomenological theory we have classified 
and analyzed the possible magnetic defects in the 
antiferromagnetic ground state and determine their 
structures. We have derived the optimal 
sizes of the defects as functions of the antiferromagnetic 
exchange, the applied magnetic field, and geometrical 
parameters of the multilayer. The calculated magnetic 
phase diagrams show the existence regions for
all types of magnetic defects. 
Experimental investigations of the 
remanent states (observed after different magnetic pre-history) 
in [Co/Pt]/Ru multilayers with wedged Co layers reveal 
a corresponding succession of different magnetic defect 
domain types.
}
\end{abstract}

%

         
\maketitle

%

Antiferromagnetically coupled 
[Co/Pt]/Ru, Co/Ir, Fe/Au, [Co/Pt]/NiO
multilayers with strong perpendicular magnetic anisotropy 
represent a new class of  synthetic magnetic materials
characterized by a cascade of field-driven 
reorientation transitions, extended 
regions of metastable states 
and specific multidomain structures 
\cite{Hellwig03,Hellwig03a,Hellwig07,Baruth06,APL07,Fu07,Hauet08}.
These spatially inhomogeneous magnetic textures 
can be separated into two fundamentally different
groups.
\textit{Regular} multidomain configurations, which
correspond to the global or local minima
of the systems, and \textit{irregular}
networks of domain walls and bands
within the antiferromagnetic ground state.
The latter are topologically stable
inclusions of ``old'' phases
trapped within the equilibrium states.
These defects display a large variability 
and their hysteretic formation strongly depends 
on the magnetic and temperature pre-history 
\cite{Hellwig03,Hellwig07,Baruth06,Fu07,Hauet08}. 

In this Letter we investigate 
the field-driven evolution of 
the regular phases and their topological defects
using a basic micromagnetic model for 
antiferromagnetically coupled multilayers 
with out-of-plane anisotropy.
The calculated magnetic 
phase diagrams show the stability limits 
of the regular equilibrium states and
indicate the regions where different
types of magnetic defects can exist.
Magnetization processes can be
analysed using these phase diagrams.
The approach provides a consistent picture 
for the formation of specific remanent states 
in antiferromagnetically coupled multilayers
and explains the physical mechanisms for the
configurational hysteresis of multidomain
states as recently observed in experiments on
[Co/Pt]/Ru \cite{Hellwig03a,
Hellwig07,Fu07,Hauet08} and
[Co/Pt]/NiO \cite{Baruth06} multilayers.
\begin{figure}
\includegraphics[width=7.5 cm]{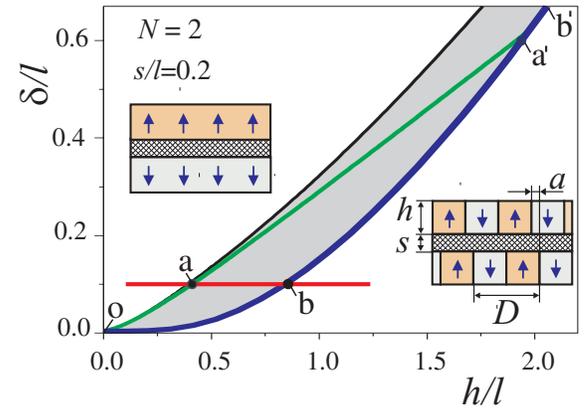}
\caption{
\label{f1}
(Color online)
The transition region between the homogeneous
antiferromagentic phase and the ferrostripes.
The first-order transition line between 
these states is shown by a thick (blue) line
$o-b-b'$.
The ferrostripes are
metastable within the shaded area.
The ferromagnetic bands
(Fig. \ref{f3} a) exist between 
 $o-a-a'$ and $o-b-b'$ lines.
Thick (red) line
$\delta/l = 0.1$ corresponds
to the horizontal axis in
Fig. \ref{f2}, 
$h_a =0.401$,
$h_b = 0.923$.}
\end{figure}

Antiferromagnetically coupled Co superlattices
investigated in 
\cite{Hellwig03,Hellwig07,Baruth06,Fu07,Hauet08}
include $N$  "ferromagnetic" blocks 
composed of $X$ bilayers [Co($h$)/M ($t$)]
antiferromagntically coupled via 
A ($s$) spacers (M = Pt, Pd, A = Ru, NiO, Ir)
with thicknesses $h$, $t$, $s$  of the
corresponding nanolayers.
Following \cite{stripes,APL07} we write
the reduced energy density ($w= W/(2 \pi M^2 X N)$
of ferromagnetic stripes as 
\begin{eqnarray}
w = 
 \frac{4 l }{D}
+\frac{\delta}{hX}\left(1- \frac{1}{N} \right)
-\frac{H q}{2 \pi M}  + w_m (D,q)\,.
\label{energy0}
\end{eqnarray}
The first term in (\ref{energy0}) is the domain wall energy,
$l= \sigma/(4 \pi M^2)$ is the \textit{characteristic}
length with  domain wall energy density $\sigma$,
the \textit{exchange length}
$\delta = J/(2\pi M^2)$
equals the ratio of  the antiferromagnetic
coupling $J>0$ and the stray-field energy,
the reduced magnetization 
$q = (d_1-d_2)/D$ is defined as ratio of
the difference between the widths of up and
down domains ($d_{1}$, $d_{2}$) and the domain
period $D= d_1 + d_2$. 
The stray field energy 
$w_m $ includes the "self" energies of individual layers
and the energies
of dipolar interactions between them.
This can be expressed 
as a set of finite integrals \cite{FTT80,stripes}.
The equilibrium domain configurations
of the stripes are derived by minimization 
of $w$ with respect to $D$ and  $q$ \cite{stripes}.
The seven control parameters of the model
($\delta/l$, $H/(4 \pi M)$, $h/l$, $s/l$, $t/l$, 
$X$, $N$) create a complex multi-dimensional
phase diagram of possible solutions.
\begin{figure}
\includegraphics[width=7 cm]{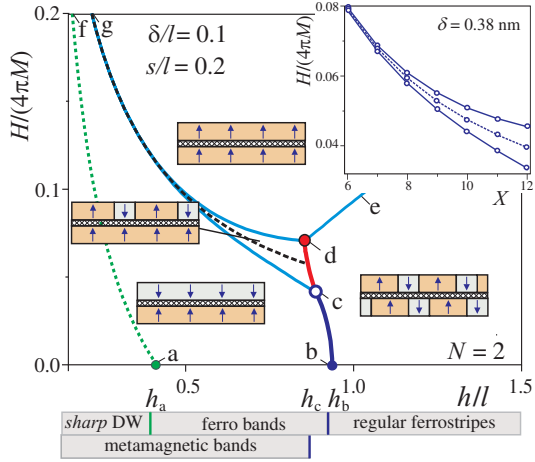}
\caption{
\label{f2}
(Color online)
The magnetic phase diagram of 
the equilibrium states
in reduced variables for layer
thickness $h/l$ and bias field
$H/(4 \pi M)$.
Metamagnetic stripes exist within
area $c-g-d$. The thick line
$c-d$ indicates the first-order
transition between metamagnetic
and shifted ferro stripes.
The shifted ferro stripes phase transforms 
discontinuously into
the antiferromagnetic
and ferromagnetic phases
along lines $b-c$ and $d-e$,
correspondingly.
The first-order lines meet
in the triple points
$c$ (0.874,0.043) and
$d$ (0.845, 0.072).
The dotted (green) line $a-f$
is the stability limit
of the ferromagnetic band
defects.
The lower panel indicates thickness
intervals for the different types
of remanent states.
The inset shows the region of the magnetic
phase diagram for [Co/Pd]/Ru multilayers
invesigated in Ref. \cite{Fu07}.
}
\end{figure}
\begin{figure}
\includegraphics[width=7 cm]{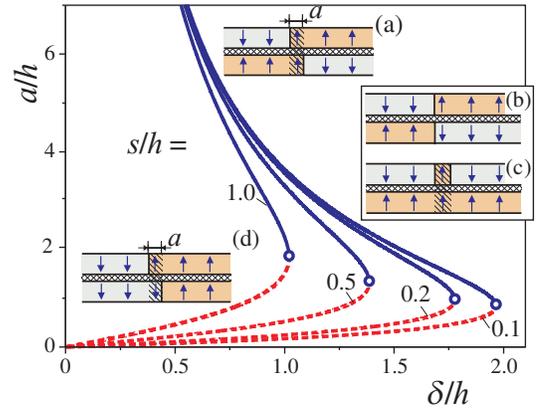}
\caption{
(Color online) 
\label{f3}
The optimal values of the
reduced width $a/h$
for \textit{ferrobands} (a)
and metastable shifted
ferrostripes (d)
(dashed lines)
as functions of the reduced 
exchange length $\delta/h$
for different thickness ratio $s/h$.
The inset shows antiferromagnetic
domains with sharp walls
(b) and
\textit{metamagnetic}
band defects (c).
}
\end{figure}
To demonstrate main features
of these solutions we consider
a simple model  for $N=2$
(Fig. \ref{f1}).
The phase diagram for weak enough 
interlayer exchange in Fig. \ref{f1}
contains the antiferromagnetic monodomain 
or ferrostripe phases. These phases are 
the possible ground states of the 
multilayers investigated in \cite{Hellwig03,Hellwig07,
Baruth06,Fu07,Hauet08}.
The antiferromagnetic
exchange coupling causes a relative
shift of domains ($a \propto \delta$) in
adjacent layers (Fig. \ref{f1}).
This lateral shift induces 
the instability of
the ferrostripe phase
at lower thickness \cite{APL07}
(Fig. \ref{f1}).
In a magnetic field the antiferromagnetic
phase transforms into the saturated state
via a specific multidomain phase.
This is similar to a metamagnetic 
phase transition in bulk antiferromagnets 
\cite{meta77}.
In an intermediate (\textit{metamagnetic}) 
phase domains arise only in 
one of the ferromagnetic blocks 
while the other remains
in the homogeneous (saturated) 
state (Fig. \ref{f2}). 
In the limit of large domains
($ D \gg L $, $L= 2Xh + 2(X-1)t + s$ is the 
multilayer thickness) the energy 
of the metamagnetic phase
can be reduced to the following form
\begin{eqnarray}
\label{MetaN2}
w = 1 - \frac{4 h X}{\pi D} 
\left[ \frac{3}{2} - \ln \left[ \frac{\pi h X}{D \cos(\pi q/2)}\right] 
- \Lambda \right] - 2 \eta q,
\end{eqnarray}
where $ \eta = H/(4 \pi M) - \delta/(2 X h)$
is the reduced strength of the field, $\tau = (t +h)/h$,
$\Lambda = \pi l/(h X) + \gamma ( X)$,
$\gamma (X) = N^{-2}\sum_{k=1}^{X-1} 
(N-k) \Upsilon_k (\tau)- \ln (X)$, 
$\Upsilon_k (\tau)= 2 \upsilon(\tau k) - \upsilon (\tau k +1)
- \upsilon (\tau k -1)$, and
$\upsilon (\omega) = \omega^2 \ln (\omega)$.
Minimization of energy (\ref{MetaN2}) yields
the following solutions \cite{FTT80}
\begin{eqnarray}
\label{stripeS}
D = 4 \pi h X/\sqrt{(\eta^*)^2 -\eta^2 }, \quad
q = (2/\pi) \arcsin \left(\eta/\eta^* \right),
\end{eqnarray}
where $\eta^* = \pi \exp (- \Lambda +1/2)$.
The upper ($H_1$) and lower ($H_2$) limiting fields
of the metamagnetic region can be written as
\begin{eqnarray}
\label{metaH}
H_{1,2} = 2 \pi M \frac{\delta}{Xh}\pm 4M 
\exp \left[-\frac{\pi l}{Xh} + \gamma(X)\right].
\end{eqnarray}
The metamagnetic transition has been
observed in [Co/Pd]/Ru
multilayers with   $X =7$ and $N = 2$,
$h = 0.4$ nm, $t = 1.8$ nm, and $s$ = 0.8 nm
\cite{Fu07}.
For this multilayer Eqs. (\ref{stripeS})
yields the domain period $D_0$ = 3.43 $\mu$m
in the center of the metamagnetic region
$H_0$ ($q =0$), and the width of this region
$\Delta H$ = 3 mT. 
For the experimental value 
$H_0 = 0.126$ T and the saturation
magnetization $4 \pi M = 1.85$ T
the equation
$H_0 = 2 \pi M (\delta/h)$ ($\eta =0$)
yields  $ \delta$ = 0.38 nm.
The calculated magnetic phase diagram, 
using this value of the exchange length,
for these [Co/Pd]/Ru systems 
(Fig. \ref{f3}, Inset) shows
a widening of the metastable
region for $X > 7$.
In the multilayers with $N \geq 4$
the metamagnetic transition occurs
first in the surface layer 
at $H \propto \delta/h$,
and then in the internal layers
at higher field $H \propto 2\delta/h$. 
This kind of two-step transition in an external field
has been observed in [Co/Pt]/Ru systems \cite{Hellwig07}.
Similarly, it also occurs in 
antiferromagnetically coupled
multilayers with in-plane magnetization \cite{PRB04}. 
%

\begin{figure}
\includegraphics[width=7.cm]{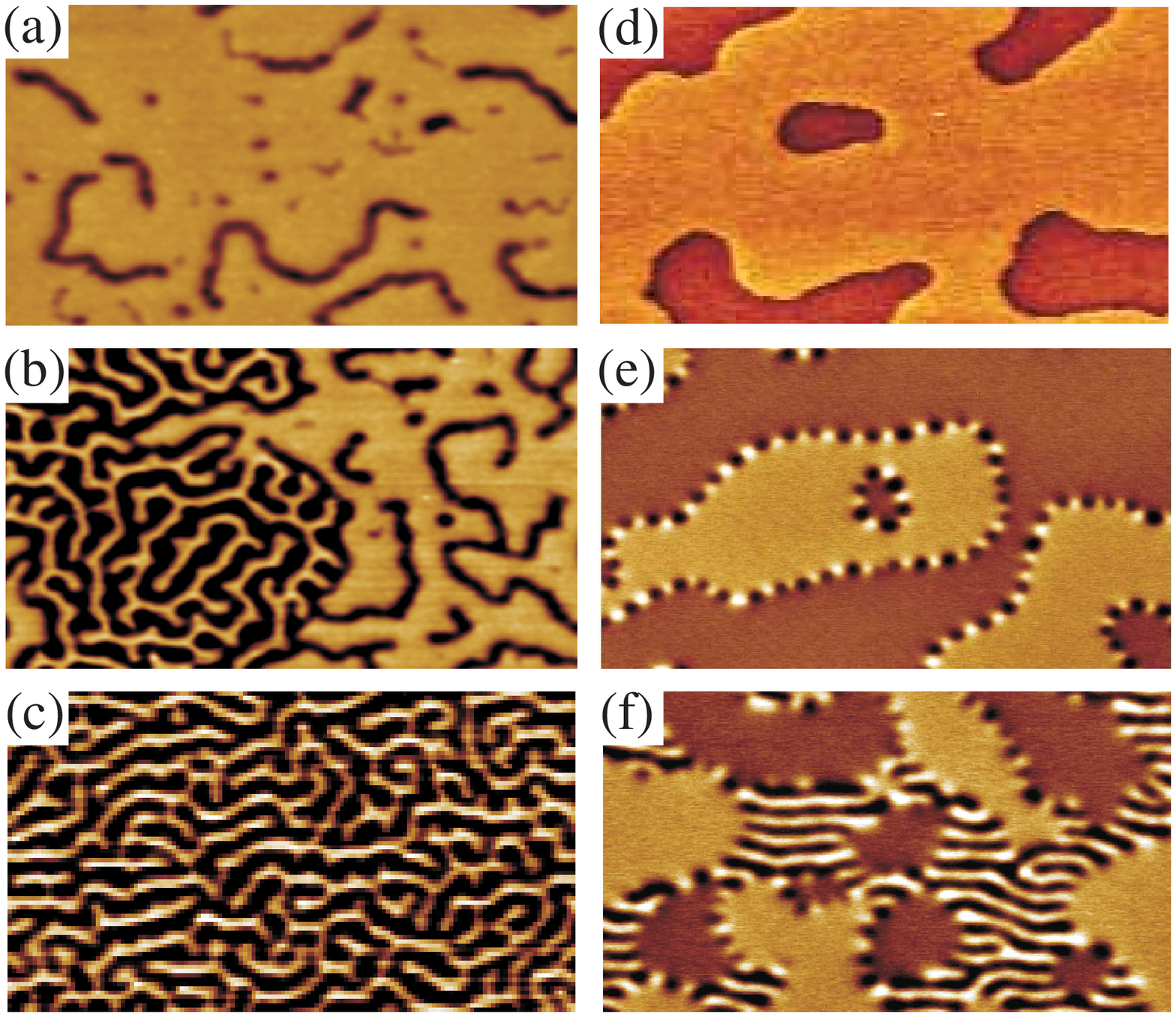}\caption{
(Color online) 
\label{f4}
Multidomain patterns 
observed in 
Pt(20)[[Co(h)Pt(0.7)]$_7$ Co(h) Ru(0.9)]$_{14}$
[Co(h)Pt(0.7)]$_8$ Pt(0.13) multilayers
with wedged Co layers after
out-of-plane saturation (a-c)
and in-plane demagnetization (d-f).
($h =$ 0.4 - 0.44  nm for a-c, e,f
and $h =$ 0.36 - 0.4 nm for d,
field of view is 7 $\mu$m $\times$ 4 $\mu$m).
}
\end{figure}

The main types of topological defects
in the antiferromagnetic ground state
include ferro bands (Figs. \ref{f3}(a),
\ref{f4} (e)), 
antiferromagnetic domains with \textit{sharp} domain
walls ( \ref{f3} (b), \ref{f4} (d)),
and metamagnetic bands 
(\ref{f3} (c), \ref{f4} (a), (b)). 
For isolated ferro bands the total energy
can be written as
\begin{eqnarray}
\label{energyBand0}
E = 2M^2 h^2 \left[ F(u, \tau) + 2 \pi (\delta/h) u \right]
\end{eqnarray}
$F(u, \tau) = 2 f (a, \tau) - f(u, \tau+1) - f (u,\tau-1)$
$f(u,\omega) = (\omega^2-u^2)\ln ( \omega^2 + u^2)
- \omega^2 \ln(\omega^2) -4 \omega u \arctan (u/\omega)$.
The equation $d E / d a = 0$ derives the optimal ferroband
widths (Fig. \ref{f3}). 
These solutions are consistent with numerical results in 
\cite{Hellwig03a,Hauet08}.
The equation $ d^2 E/d u^2 =0$ yields 
the critical value of the band width
$a_{cr} =\sqrt{s^2 + 2 s h +h^2/2}$.
Substituting this solution into equation $d E / d a = 0$ 
we obtain the equation for the lability line
($o-a-a'$ in Fig. \ref{f1}).
The equation $H = 2 \pi M \delta/h_a$ defines the stability
limit of the ferroband in the magnetic phase diagram 
(line $a-f$ in Fig. \ref{f2}).
Usually the ferrobands split into domains
with up and down magnetization creating
exotic patterns, named ``tiger-tails'' \cite{Hellwig03a}.
The ferrobands exist within the lability
region of the ferrostripe phase (Fig. \ref{f1}).
They collapse into sharp domain walls 
at the lability $o-a-a'$ and transform
into the ferrostripes by unlimited expansion
of "tiger-tail" patterns at the transition
line $o-b-b'$ as  recently observed in
[Co/Pt]/Ru multilayers  (Fig. \ref{f4} (e),(f)
and \cite{Hellwig07}).
The phase diagrams of solutions (Figs. \ref{f1}, \ref{f2})
explain the evolution of multidomain states observed
in \cite{Hellwig03,Hellwig07,Fu07,Hauet08}
and the formation of different remanent states
in antiferromagnetically coupled multilayers
 \cite{Hellwig03a,Hellwig07,Baruth06,Fu07,Hauet08}.
Usually after out-of-plane saturation 
magnetic patterns at zero field
consist of remnants of the metamagnetic
domains ($h < h_c$ in Fig. \ref{f3} and
Fig. \ref{f4}, (a)), a mixture
of metamagnetic remnants and ferrostripes
($h_c < h < h_b$, Fig. \ref{f4} (b)), and
regular ferrostripe textures ($h > h_b$,
Fig. \ref{f4} (c))
($h_a$, $h_b$, $h_c$ are defined on the
horizontal axis in Fig. \ref{f2}).
In-plane demagnetization yields a succession
of remanent states consisting of
antiferromagnetic domains with sharp
domain walls ($h < h_a$, Fig. \ref{f4}, (d))
or a network of ferrobands 
($h_a< h < h_b$, Fig. \ref{f4}, (e)).
These domains
act as nucleation regions of ferrostripes
within the antiferromagnetic state
(Fig. \ref{f4} (f)). 
For $h > h_b$
regular ferrostripes create the ground
state of the  multilayer (Fig. \ref{f4}, (c)).
 
In conclusion, we have developed
a micromagnetic theory for
regular magnetic phases and topological
defects arising in perpendicular
antiferromagnetically coupled multilayers
and explain the formation of different remanent
states as consequence of topologically 
stable defects with different geometry.

 
\begin{acknowledgments}
The authors thank  T. Hauet, J. McCord and
R. Sch{\"a}fer for helpful discussions.
Work supported by DFG through SPP1239 (project A08).
N.S.K. and A.N.B.
\ thank H.\ Eschrig for support and
hospitality at IFW Dresden. 
\end{acknowledgments}

\end{document}